**Title**: Correcting for T$_1$ bias in Magnetization Transfer Saturation (MT$_{sat}$) Maps Using Sparse-MP2RAGE

**Authors:**

Christopher D. Rowley[1,2,3], Mark C. Nelson[1], Jennifer S.W. Campbell[1], Ilana R. Leppert[1], G. Bruce Pike[4], Christine L. Tardif[1,2,5]

[1] McConnell Brain Imaging Centre, Montreal Neurological Institute and Hospital, McGill University, Montreal, QC, Canada H3A 2B4

[2] Department of Neurology and Neurosurgery, McGill University, Montreal, QC, Canada H3A 2B4

[3] Department of Physics and Astronomy, McMaster University, Hamilton, ON, Canada L8S 4L8

[4] Hotchkiss Brain Institute and Departments of Radiology and Clinical Neuroscience, University of Calgary, Calgary, Canada T2N 4N1

[5] Department of Biomedical Engineering, McGill University, Montreal, QC, Canada H3A 2B4

**Corresponding Author:**

Christine Tardif, Department of Biomedical Engineering, McGill University, Montreal, QC, Canada H3A 2B4 christine.tardif@mcgill.ca

Christopher Rowley, Department of Physics and Astronomy, McMaster University, Hamilton, ON, Canada. L8S 4L8

**Abstract:**

**Purpose:**

Magnetization transfer saturation (MT$_{sat}$) mapping is commonly used to examine the macromolecular content of brain tissue. This study compared variable flip angle (VFA) T$_1$ mapping against compressed sensing (cs)MP2RAGE T$_1$ mapping for accelerating MT$_{sat}$ imaging.

**Methods:** VFA, MP2RAGE and csMP2RAGE were compared against inversion recovery (IR) T$_1$ in a phantom at 3 Tesla. The same 1 mm VFA, MP2RAGE and csMP2RAGE protocols were acquired in four healthy subjects to compare the resulting T$_1$ and MT$_{sat}$. Bloch-McConnell simulations were used to investigate differences between the phantom and *in vivo* T$_1$ results. Finally, ten healthy controls were imaged twice with the csMP2RAGE MT$_{sat}$ protocol to quantify repeatability.

**Results:** The MP2RAGE and csMP2RAGE protocols were 13.7% and 32.4% faster than the VFA protocol, respectively. All approaches provided accurate T$_1$ values (<5% difference) in the phantom, but the accuracy of the T$_1$ times was more impacted by differences in T$_2$ for VFA than for MP2RAGE. *In vivo*, VFA generated longer T$_1$ times than MP2RAGE and csMP2RAGE. Simulations suggest that the bias in the T$_1$ values between VFA and IR-based approaches (MP2RAGE and IR) could be explained by the MT-effects from the inversion pulse. In the test-retest experiment, we found that the csMP2RAGE has a minimum detectable change of 3% for T$_1$ mapping and 7.9% for MT$_{sat}$ imaging.

**Conclusions:**

We demonstrated that csMP2RAGE can be used in place of VFA T$_1$ mapping in an MT$_{sat}$ protocol. Furthermore, a shorter scan time and high repeatability can be achieved using the csMP2RAGE sequence.

**Introduction:**

Magnetization transfer saturation ($MT_{sat}$) is a semi-quantitative method that is used commonly as a biomarker for tissue myelin content (Helms et al., 2010, 2008b). The technique uses magnetization transfer (MT) contrast, which is generated using an off-resonance radiofrequency pulse that partially saturates the bound protons in the macromolecular pool. This decreased, or partially saturated, state of magnetization is transferred to the water pool and presents itself as a decrease in MRI-observable signal. While the MT ratio (MTR) is a more simplistic method to investigate these effects, $MT_{sat}$ aims to quantify this saturation while accounting for counteracting effects of $T_1$ relaxation. The calculation requires an MT-weighted image and knowledge of the observed longitudinal relaxation time ($T_{1obs}$, shortened herein to $T_1$) and apparent equilibrium magnetization ($M_0$).

Typically, the $M_0$ and $T_1$ maps for this purpose are generated using a variable flip angle (VFA) experiment (Christensen et al., 1974; Helms et al., 2008b; Venkatesan et al., 1998; Weiskopf et al., 2013). VFA $T_1$ mapping involves using a spoiled gradient-recalled echo (SPGR) or fast low angle shot (FLASH) sequence to acquire two or more images with different $T_1$-weightings, where the timing and flip angles of each image can be optimized for signal-to-noise in the final map (Helms et al., 2011). $T_1$ values can then be obtained through fitting (Gupta, 1977) or solving signal equations (Helms et al., 2008a). VFA $T_1$ mapping has been widely adopted due to its ease of use (uses product sequences, simple calculation), and for its ability to generate high-resolution 3D maps much faster than more conventional approaches such as inversion recovery (IR) (see (Kingsley, 1999) for a list of seminal papers on $T_1$ mapping). It also uses the same base pulse sequence that is used for the MT-weighted image, removing sequence-dependent receiver gain differences, matches geometric distortions and has the same $T_2^*$ contributions due to matching echo times.

However, VFA $T_1$ mapping has been shown to provide $T_1$ times up to 20% longer compared to the gold standard IR approach (Stikov et al., 2015b; Tsialios et al., 2017). Post-processing of VFA $T_1$ maps can correct for flip angle inaccuracies caused by $B_1^+$ inhomogeneities ($\Delta B_1^+$) (Venkatesan et al., 1998), and for incomplete spoiling of the transverse magnetization between excitation pulses (Baudrexel et al., 2018; Preibisch and Deichmann, 2009). This removes spatial variations that correlate with $\Delta B_1^+$ and typically reduces VFA $T_1$ times in the brain (Baudrexel et al., 2018), but differences remain (Stikov et al., 2015a). This increase in $T_1$ will result in decreased $MT_{sat}$ values due to the inverse relationship between the two metrics.

$MT_{sat}$ values are also impacted by $\Delta B_1^+$. It is possible to remove these effects using an empirically derived factor that corrects $\Delta B_1^+$ induced errors in the $MT_{sat}$ map from the MT-weighted image and the VFA-derived $T_1$ map (Helms, 2015; Weiskopf et al., 2013). A recent framework has shown that it is possible to correct for the $\Delta B_1^+$ effects in the saturation and excitation pulses separately (Rowley et al., 2021), which would permit the use of other $T_1$ mapping techniques for generating $\Delta B_1^+$ corrected $MT_{sat}$ maps. To this end, this study explores the use of MP2RAGE $T_1$ mapping in an $MT_{sat}$ protocol.

The MP2RAGE sequence uses two readout blocks following an inversion pulse, where each readout block typically acquires one plane of k-space, to generate two images with different inversion times. The two acquired images are combined with a lookup table derived from Bloch simulations to extract $M_0$ and $T_1$ maps (Marques et al., 2010). MP2RAGE $T_1$ values have been shown to be in strong agreement with gold standard IR $T_1$ mapping (Tsialios et al., 2017). This sequence has a decreased dependence on $\Delta B_1^+$, and no motion differences between the images due to their simultaneous acquisition. Despite the longer TR necessary with the inclusion of an inversion pulse, the shorter echo-spacing in the readout allows the MP2RAGE sequence to be a more efficient at $T_1$ mapping than the VFA approach. Compressed sensing approaches have been implemented in quantitative MRI protocols previously, including MTsat using VFA imaging available on Philips scanners (Berg et al., 2022). Here, we explored the use of the Siemens compressed sensing MP2RAGE (csMP2RAGE) (Mussard et al., 2020) product sequence for further accelerating an $MT_{sat}$ protocol.

The overall aim of this work is to acquire fast and accurate $MT_{sat}$ maps using csMP2AGE instead of VFA to remove $T_1$ effects. This study compares the accuracy and repeatability of $M_0$ and $T_1$ values generated using 2D IR, as well as 3D VFA, MP2RAGE and a csMP2RAGE sequences in a calibrated phantom. Next, we investigate how the three accelerated 3D $T_1$ mapping methods compare *in-vivo* for generating $MT_{sat}$ maps. We use Bloch-McConnell numerical simulations of the sequences to provide insight into the source of the differences in the observed $T_1$ times. Finally, we present repeatability results from a test-retest imaging experiment in ten healthy controls using the csMP2RAGE sequence for accelerated $MT_{sat}$ imaging.

## Methods:

### Data acquisition:

This study was approved by our institutional ethics committee. MR images were acquired using a 3T-PrismaFit scanner (Siemens, Germany) with a 32-channel receive coil. The imaging parameters are included in **Table 1**.

To assess the accuracy of the methods, the ISMRM/NIST system phantom (Stupic et al., 2021) (Premium System Phantom, CaliberMRI, Boulder, Colorado) was imaged. 2D spin echo IR images were acquired over the $T_1$ ($NiCl_2$) and $T_2$ ($MnCl_2$) plates, where each consists of 14 spheres with varying $T_1$ and $T_2$ values (listed in **Table 2**). Only the spheres with $T_1$ values greater than 300 ms were used in the analysis. Both arrays were used to have more samples within the range of healthy brain tissue (700-1600 ms), and to observe potential biases from different $T_2$ values. MT data was not acquired in the NIST phantom as the solution in the spheres are aqueous and have no macromolecular-associated MT effects.

For comparing pulse sequences, four healthy adults (aged 29-35 years, three female) were imaged using the same hardware and protocol without the IR $T_1$ mapping. For the test-retest comparison, ten different healthy adults (aged 20-40 years, four female) were imaged using the identical hardware setup but only with the csMP2RAGE $MT_{sat}$ protocol.

All sequences used the Siemens *prescan normalize* option to generate $M_0$ maps with minimal receive bias. This normalization process incorporates a calibration scan obtained from the body receive coil to correct the sensitivity profile of the head coil. A gain factor of 2.5 was applied to $M_0$ maps from the MP2RAGE protocols to match the receiver gain factor of the MT-weighted image. $B_1^+$ maps were acquired using a fast turboFLASH sequence (Chung et al., 2010).

For *in-vivo* imaging, a $T_1$-weighted MPRAGE image was acquired as an anatomical reference with the following parameters: 1 mm isotropic, TR = 2300 ms, TE = 2.98 ms, flip angle = 9 deg, TI = 900 ms, FOV = 256x256x192, and a GRAPPA factor of 2 with 32 reference lines. This was done to allow all images of interest to be registered to a common space.

**Data fitting:**

Image calculations were done in MATLAB 2021b (www.mathworks.com). Spin echo IR $T_1$ values were fit using the qMRlab toolbox (Karakuzu et al., 2020) using a shortened TR signal equation (Barral et al., 2010).

MP2RAGE-based $M_0$ and $T_1$ maps were fit using a lookup table approach that incorporates $\Delta B_1^+$. This was done by modifying the original MP2RAGE code (https://github.com/JosePMarques/MP2RAGE-related-scripts) to extract a $\Delta B_1^+$ corrected $M_0$ map from the lookup table, in addition to the $T_1$ map.

The hMRI toolbox was used to calculate the $T_1$ values from the VFA data (Tabelow et al., 2019) without the small tip angle approximation (Edwards et al., 2021). For the NIST phantom data, spoiling correction was not applied to VFA metrics due to the wide range of $T_2$ values. For in-vivo data, the $T_1$ values were corrected for spoiling imperfections (Preibisch and Deichmann, 2009), which was implemented using code from hMRI toolbox for a $T_2$ value of 80 ms. This is done by fitting the VFA data to extended phase graph simulations of the sequence (Malik et al., 2018) and was included in our model-based corrected VFA values. The corrected $M_0$ values were obtained from each of the two images by inputting the $\Delta B_1^+$ corrected $T_1$ values into the FLASH signal equation (equation 1) and then averaging.

$$S = M_0 \sin(\alpha\, B_1^+)\, \frac{1 - exp(-TR/T_1)}{1 - cos(\alpha\, B_1^+) exp(-TR/T_1)} \quad (1)$$

MT$_{sat}$ maps were computed as in (Helms et al., 2008b) using:

$$MT_{sat,uncor} = \left(\frac{M_0 \cdot \alpha}{S_{MTw}} - 1\right) \cdot \frac{TR}{T_1} - \frac{\alpha^2}{2} \quad (2)$$

Where $\alpha$ is the excitation flip angle in the MT-weighted image, $S_{MTw}$ is the MT-weighted signal, and $TR$ is the repetition time of the MT-weighted sequence. For the MT$_{sat}$ values derived from MP2RAGE $T_1$ maps, a model-based correction was used to correct for $\Delta B_1^+$ (Rowley et al., 2021). This same approach was used for the model-based VFA MT$_{sat}$ values. To compare with what is most frequently found in the literature, a VFA MT$_{sat}$ map was generated using a previously generated empirical correction factor (Helms, 2015; Weiskopf et al., 2013) defined as:

$$MT_{sat,cor} = MT_{sat,uncorr} \left(\frac{1 - 0.4}{1 - 0.4 \cdot \Delta B_1^+}\right) \quad (3)$$

**Regions of Interest (ROI):**

Since the 3D and 2D data were acquired with different resolutions over different regions of the phantom, ROIs were manually delineated for the 2D and 3D data (**Figure 1**). The denoised MP2RAGE UNI image was used for manually delineating a 3D mask for the spheres of interest in the NIST phantom in ITK-snap (Yushkevich et al., 2016). 2D masks were generated from the TI = 1000 ms inversion recovery images of each sphere array. Masks were drawn in the image from the first session, with each sphere as a separate label and registered using ANTS (Avants et al., 2008) to align the mask to the subsequent imaging sessions. To avoid edge artifacts, the registered 3D masks for each session were eroded using a sphere with a radius of four voxels in MATLAB, and the 2D masks were eroded using a disk with a radius of two voxels.

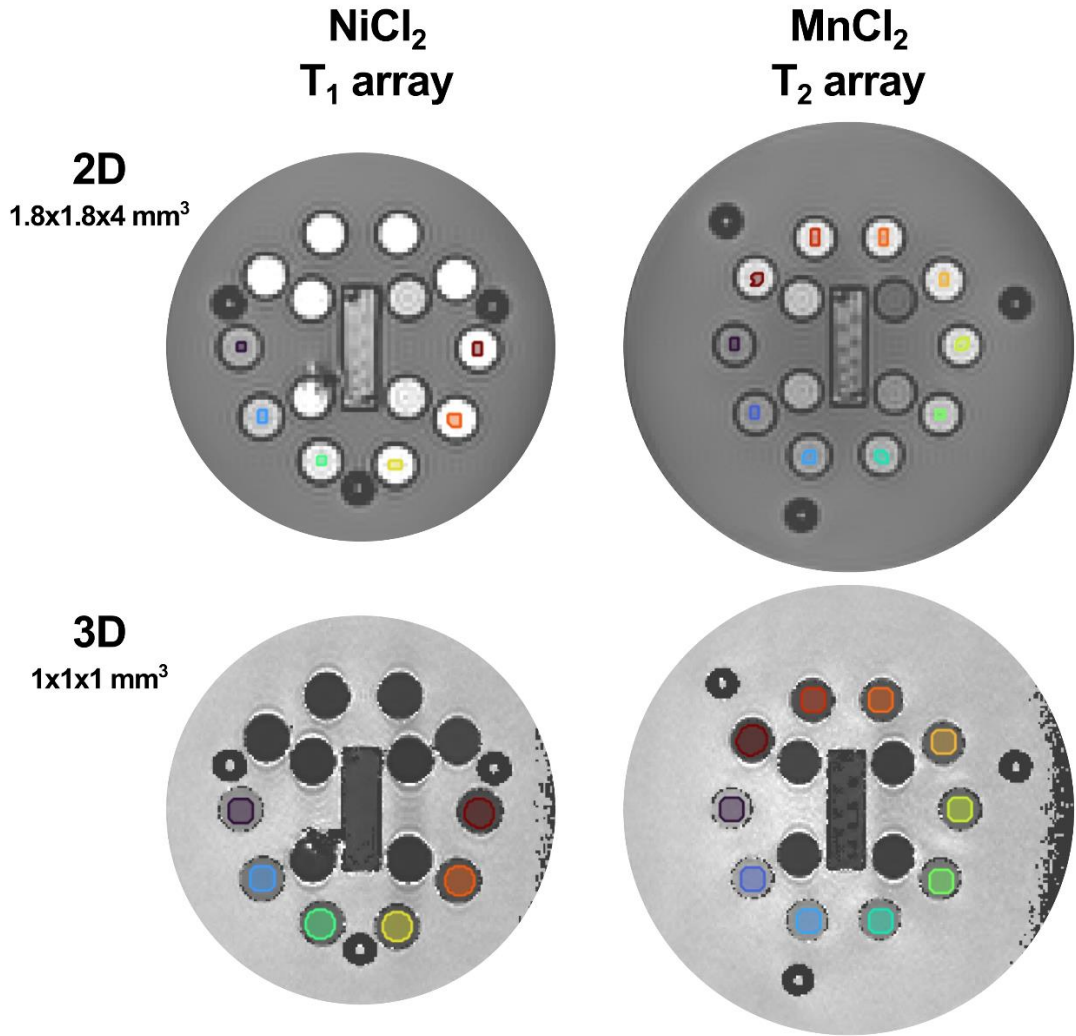

**Figure 1**: NIST phantom arrays with the ROIs overlaid. Separate ROIs were used for the 2D and 3D images for the $T_1$ and $T_2$ sphere plate arrays. ROIs were manually drawn and eroded to avoid edge artifacts.

For the human data, all images were registered to the $T_1$-weighted MPRAGE image, which was used as input to the FreeSurfer (v7.2) (Fischl, 2012) *recon-all* pipeline with manual edits to remove dura matter. An ROI analysis was performed utilizing the *wmparc.mgz* segmentation from FreeSurfer for its volumetric white matter (WM) and subcortical gray matter (GM) labels (Desikan et al., 2006). For cortical GM ROIs, the calculated metrics were sampled at the mid-depth cortical surface and averaged over the FreeSurfer *aparc.a2009s* surface ROIs (Destrieux et al., 2010). ROIs were removed with missing data, leaving 89 volumetric ROIs and 75 surface ROIs per subject.

**Statistical Analysis:**

Metrics were compared using a framework that was recently presented for quantitative imaging of the spinal cord (Lévy et al., 2018). Repeatability (test-retest) was examined in the NIST phantom and ten healthy controls. This involved computing the intraclass correlation coefficient (ICC) and the minimum detectable change (MDC). ICC is the ratio of the inter-subject variance to the total observed variance. This provides insight into the amount of observed variation that is due to genuine between-subject differences compared to variability from the imaging and processing. The MDC defines the minimum necessary difference in a metric needed to report a statistically significant change and is used to assess reliability and the precision of the approach. The MDC is calculated from repeat measurements and is defined as:

$$MDC = \sqrt{2} \cdot 1.96 \cdot \sigma_{within} \qquad (4)$$

where $\sigma_{within}$ is the within subject standard deviation. To assess differences between metrics calculated from different pulse sequences, Bland Altman plots (Altman and Bland, 1983; Giavarina, 2015) were used.

Simulations:

To investigate the MT effect of the different $T_1$ mapping protocols, simulations of VFA, IR and MP2RAGE sequences were run using code developed for optimizing inhomogeneous MT imaging (Rowley et al., 2023), while following standards set by previous studies. The following tissue parameters were used: $R_{1B} = R_{1A} = R_{1obs} = 1.176$ s$^{-1}$; $M_{0A} = 1$; $k_f = 4.45$ s$^{-1}$; $k_r = 28.34$ s$^{-1}$; $R_{2A} = 12.3$ s$^{-1}$. $R_1 = 1/T_1$, subscripts *A* and *B* respectively correspond to free and bound proton pools, subscript *obs* is the observed $R_1$ which we will maintain for the simulation section, *k* is the exchange rate for forward (f) or reverse (r) exchange, and $R_{2A}$ is the transverse relaxation rate of the free protons. There are inconsistencies in the MT literature as to the relationship between $R_{1B}$, $R_{1A}$, and $R_{1obs}$. It is often reported that the value of $R_{1B}$ is poorly fit from traditional qMT experiments (Henkelman et al., 1993), thus we opted to use the convention used by Gloor et al. to set them equal (Gloor et al., 2008). Other studies have fixed these variables in simulations to values that are close to each other (Kim et al., 2014; Teixeira et al., 2019). The relationship previously used by Sled et al., (Henkelman et al., 1993; Sled and Pike, 2001), was derived from the SPGR signal equation and presented an unfair comparison across sequences.

To model the impact of the on-resonance excitation pulses on the bound pool, the absorption of the macromolecular lineshape (typically denoted by variable g($\Delta$), where $\Delta$ is the offset frequency) was fixed to 1.4e-5 s$^{-1}$ as in (Gloor et al., 2008; Teixeira et al., 2019). This saturated the bound pool to 88% of its initial value following an inversion pulse as was explicitly set in previous simulations (Kim et al., 2014). Given that methods exist for modeling spoiling imperfections, and therefore could be corrected in post-processing steps, we assumed the signal was perfectly spoiled after each excitation pulse. To investigate the potential impact of MT effects, all tissue parameters were fixed, and the bound pool fraction ($M_{0B}$) and $R_{1obs}$ varied to examine the MT-related impact of the excitation pulses on the resulting $T_1$ values. This may overestimate the impact that would be observed in humans *in-vivo*, as tissue parameters are often linked. To investigate the impact that would more likely be observed in healthy tissue, additional simulations were run using linear models that were built from qMT results reported previously (Sled and Pike, 2001) to modify the variables $M_{0B}$, $k_f$, $R_{1A}$, $T_{2A}$, and $T_{2B}$ based on the input $R_{1obs}$ (Appendix equations A1-A6).

MATLAB code to calculate the corrected maps and run the simulations are available at https://github.com/TardifLab/ MTsatMP2RAGE.

## Results:

Our main motivation to use csMP2RAGE over VFA for $MT_{sat}$ mapping is to shorten the acquisition protocol. The full $MT_{sat}$ protocol including $B_1^+$ mapping using csMP2RAGE or MP2RAGE were 32.4% and 13.7% faster than the VFA approach, respectively (see **Table 1**).

All Protocols Provided Accurate $T_1$ Times in the NIST Phantom

The second objective was to verify the accuracy of these $T_1$ mapping techniques against reference values and the gold standard IR approach. The NIST phantom provided insight into the accuracy of the $T_1$ mapping methods in the absence of MT effects, as well as the impact of different $T_2$ values which may impact spoiling efforts.

MP2RAGE and VFA protocols that fit $T_1$ based on two measurements, require the sequence timing to be optimized for fitting over a specific range of $T_1$ values. Outside this range, the fit results can be poor, and this is made apparent in Figure 2A and 2B where the $T_1$ values below 475 ms are poorly fit in the MP2RAGE scans. From Figures 2A, 2B and 2F, we observe that VFA tends to fit longer $T_1$ times than IR and MP2RAGE scans. As the two arrays have different $T_2$ times for similar $T_1$ ranges, Figures 2C, 2D and 2E provide insight into the $T_2$ dependence of the fitting compared to the IR approach. The clear separation of the VFA points from the two arrays suggests that this $T_1$ mapping method is additionally sensitive to parameters outside of $T_1$. The MP2RAGE-based values demonstrated increased differences with increasing $T_1$ values but presented good agreement over the range of $T_1$ times of the sphere arrays. The repeatability of the metrics was investigated over four sessions and the results are presented in Figure 2G. To better represent the brain, only ROIs with $T_1$ values above 600 ms were used in this calculation. Despite the offsets of the VFA $T_1$ values, it had a lower MDC (4.2%) compared to IR (5.8%), MP2RAGE (6.7%) and csMP2RAGE (6.2%). All metrics presented high ICC values ($\geq 0.998$).

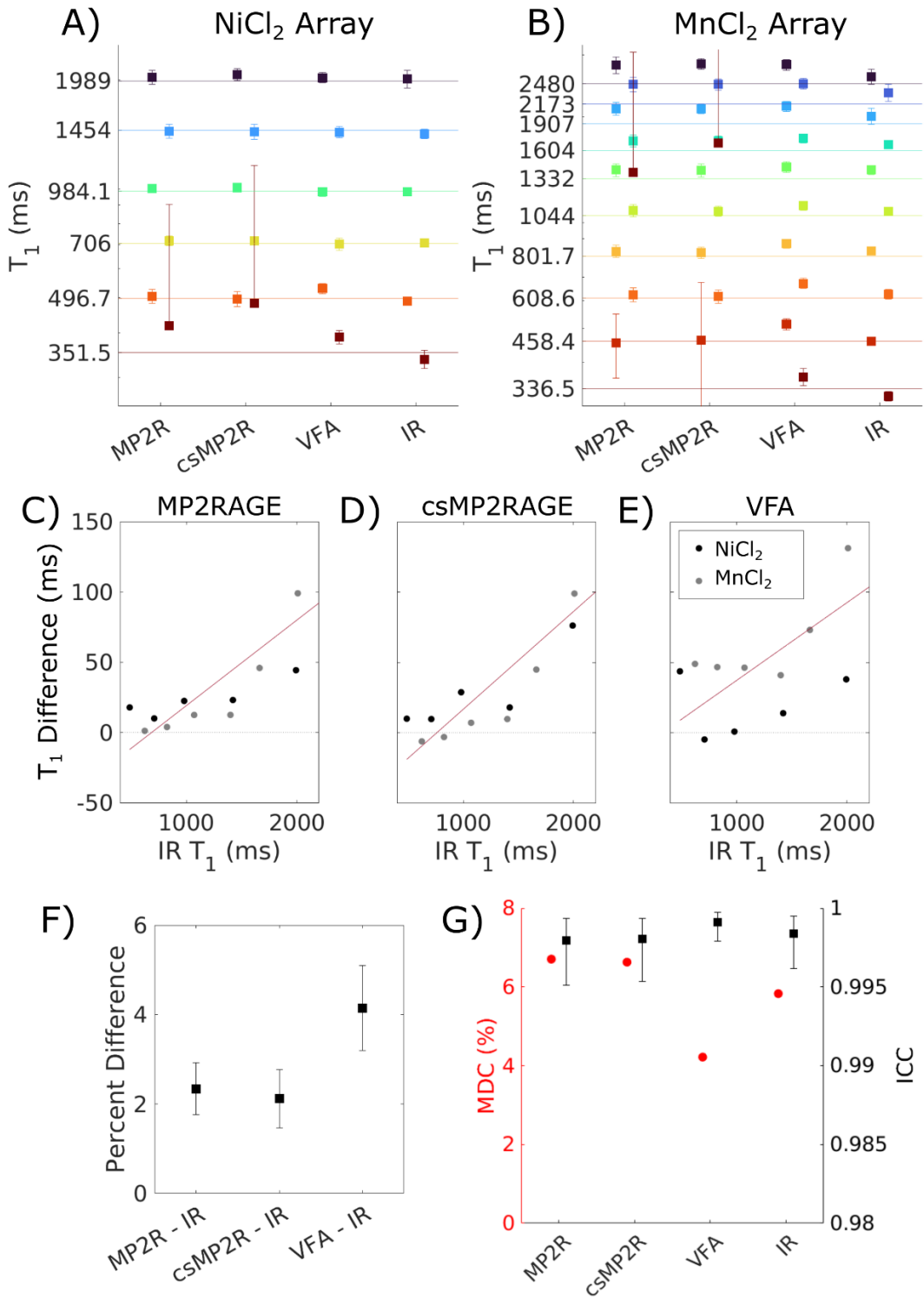

**Figure 2: A, B)** ROI analysis of NIST phantom spheres from the $T_1$ and $T_2$ arrays. Y-axis is log scaled. Data points are the average of the four sessions, horizontal lines mark the reference value for the ROI and the error bars are the standard deviations, colour is matched to ROIs in Figure 1. **C-E)** Shows the difference between accelerated 3D $T_1$ mapping methods compared to the gold standard 2D IR method. **F)** Average percent difference compared to IR, across all ROIs, averaged over four sessions. **G)** Repeatability analysis from $T_1$ values acquired over four sessions. VFA provided the most consistent

values in the NIST phantom, presenting the lowest minimum detectable change, and highest intraclass correlation coefficient.

Longer VFA $T_1$ Times than MP2RAGE Results in Lower $MT_{sat}$ *In Vivo*

Representative maps from a single subject are displayed in **Figure 3**. Visually, there is strong agreement between the $M_0$ maps. VFA produced longer $T_1$ times, resulting in lower $MT_{sat}$ values compared to the MP2RAGE-based maps. Minimal additional blurring is observed in the csMP2RAGE map despite the increased acceleration factor.

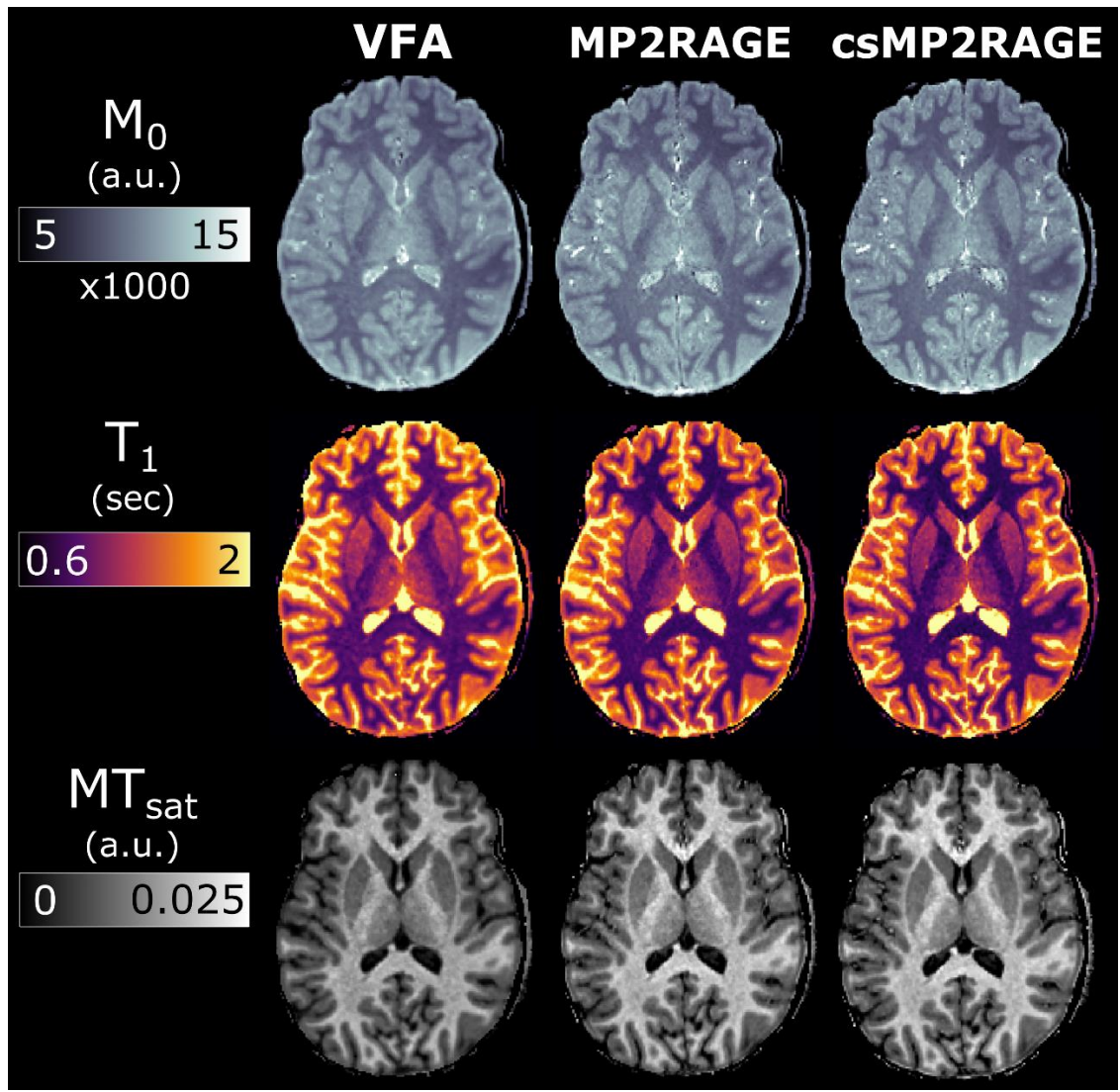

**Figure 3:** Representative axial slice from a single subject derived from the different $T_1$ mapping protocols. All maps presented here were corrected for $\Delta B_1^+$, and the VFA maps were corrected for incomplete spoiling.

The Bland-Altman plots comparing the $T_1$ and $MT_{sat}$ maps from the three $T_1$ mapping protocols are presented in **Figure 4**. The VFA approach had a shift of +73 ms in the $T_1$ values compared to MP2RAGE (Fig. 4A), resulting in a -10.82% offset in the $MT_{sat}$ values (Fig 4D). The trend was similar when comparing VFA to csMP2RAGE with offsets of +87 ms and -11.78% (Fig. 4B and E). Strong agreement was observed between the MP2RAGE and csMP2RAGE sequences with a calculated difference of -14 ms and +0.965 % for $T_1$ and $MT_{sat}$ respectively (Fig. 4C and 4F). The difference between metrics appears to have minimal dependance on $T_1$ in this range of $T_1$ values.

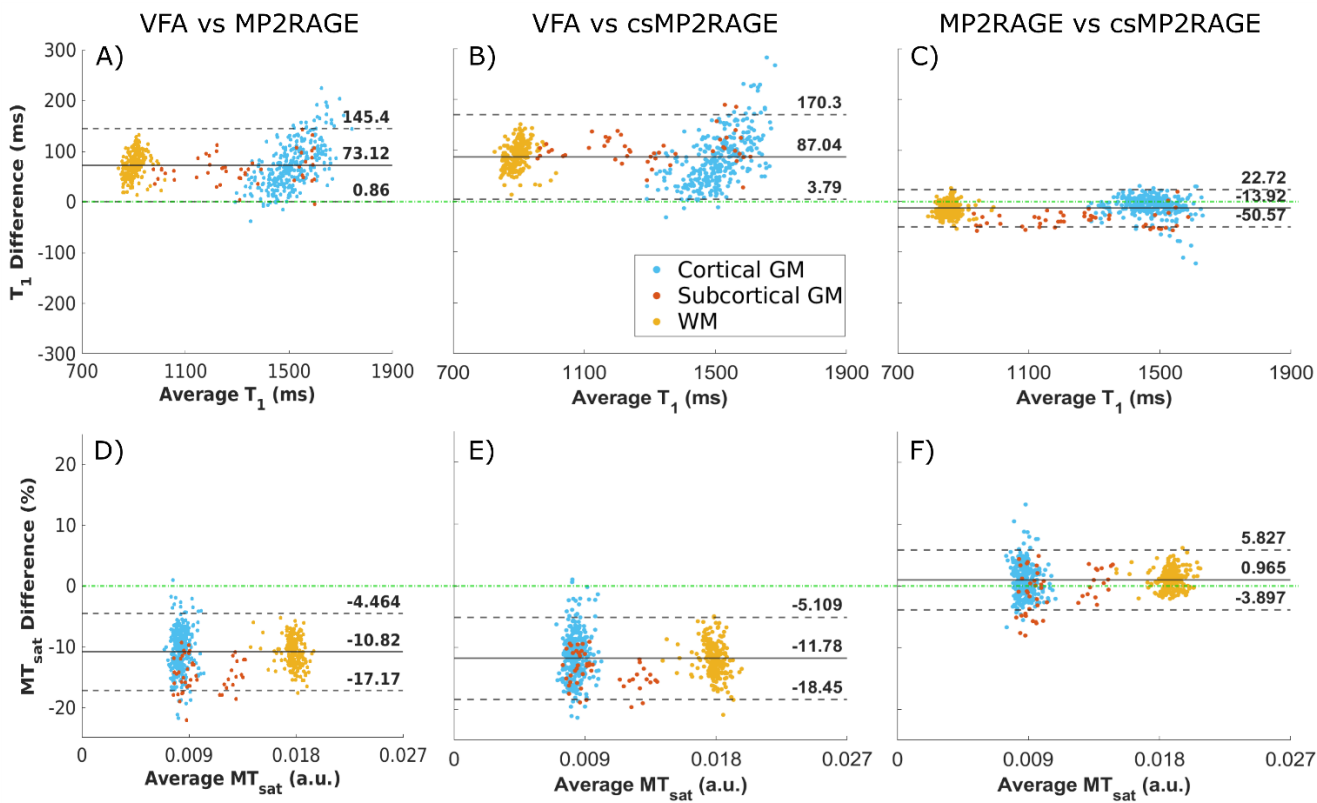

**Figure 4**: Bland Altman plots to compare $T_1$ and $MT_{sat}$ values between methods in four subjects. Data points represent the ROI averaged values from four healthy adults. The dashed green line indicates no difference between the metrics. The solid black line indicates the average difference, with the dashed lines representing the 95% confidence interval.

VFA is More Impacted By $\Delta B_1^{\pm}$

Reducing scanner-related image variations decreases the MDC of a metric. **Figure 5** highlights the impact of $\Delta B_1^+$ correction on the different maps. There was more variation in the MP2RAGE values caused by $\Delta B_1^+$ than was present between the MP2RAGE and csMP2RAGE from Figure 4. The large spread of values in Figure 5B indicates that $\Delta B_1^+$ correction is necessary for VFA $T_1$ values. Figure 5D highlights the partial cancellation of $\Delta B_1^+$ effects inherent in the $MT_{sat}$ calculation, where ±30% differences in $T_1$ reduced to -16% to +6% differences in $MT_{sat}$.

Finally, we compared the commonly employed empirical correction for $MT_{sat}$ against a model-based approach that included corrections for incomplete spoiling in the $T_1$ map, which is displayed in Figure 5E. While there is a slight bias towards lower $MT_{sat}$ values (-3.6%) when using the empirical correction, there is a strong agreement between the values after accounting for the bias. This suggests that either approach could be used to achieve a strong decrease in scanner-induced variance in $MT_{sat}$ values.

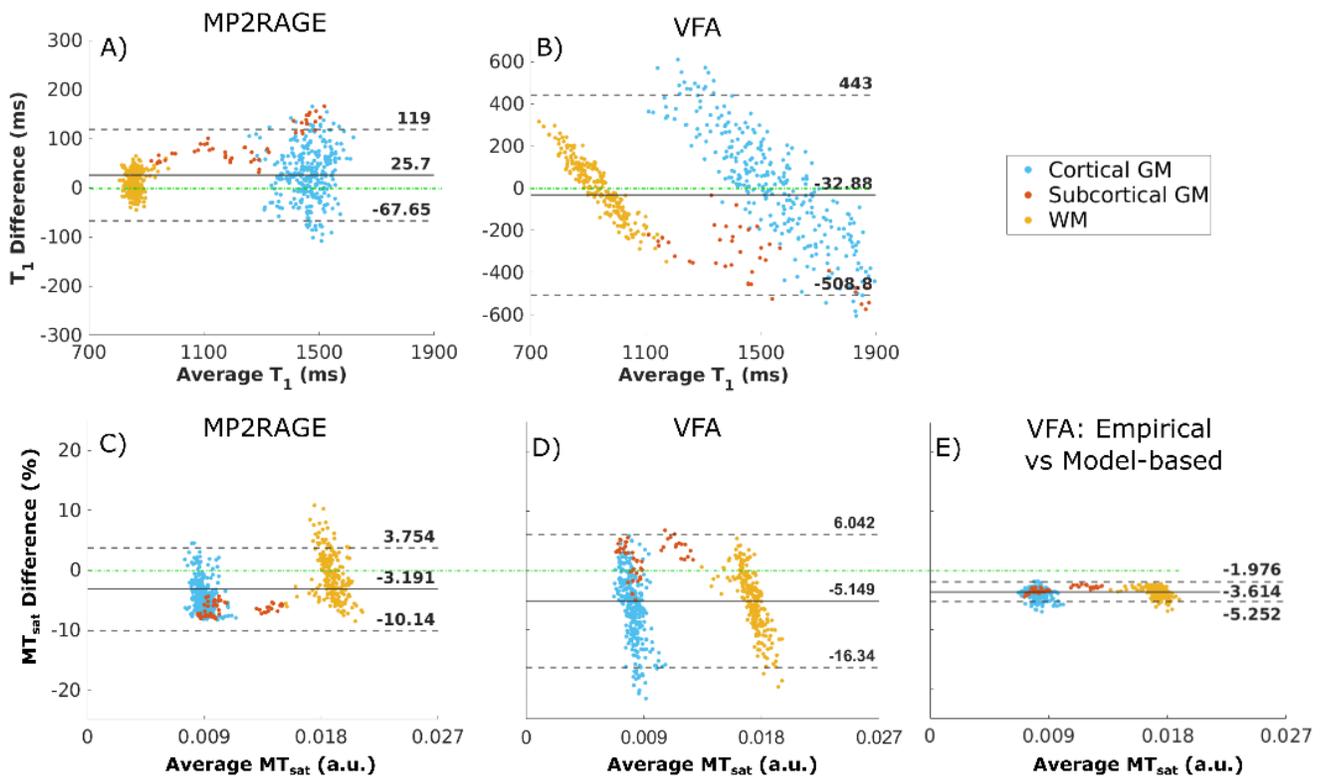

**Figure 5:** Bland Altman plots compare the results obtained within a method when $\Delta B_1^+$ is corrected for or ignored. This provides an idea of the results based on different processing streams. Note that Plot A and B use different y-axis scaling. Data points represent the ROI averaged values from four healthy adults, the solid black line indicates the average difference, with the dashed lines providing the 95% confidence interval.

Protocol-Dependent MT effects on $T_1$ Mapping

Modified Bloch-McConnell simulations were used to investigate if MT effects could describe the varying levels of agreement in the $T_1$ times obtained from the NIST phantom and *in vivo* acquisitions and are presented in **Figure 6**. Plots 6A-6E assume that only $T_{1,obs}$ and $M_{0B}$ are changing. Plots 6A-6C highlight the impact that changes in $M_{0B}$ have on the calculated $T_{1,obs}$ values. The IR and MP2RAGE have larger deviations with increasing $M_{0B}$ than the VFA protocol due to the 180-degree inversion pulse. IR is considered the gold standard for $T_1$ mapping, and simulations suggest that MP2RAGE can provide values in strong agreement with IR as seen in Figure 6D. MP2RAGE overestimated $T_{1,obs}$ on average by 4% with minimal influence from changes to $M_{0B}$. VFA values showed little difference compare to IR with $M_{0B} = 0$, which matches the results from the NIST phantom. Large deviations in output $T_{1,obs}$ (>20%) are noted however with increasing $M_{0B}$ as observed in Figure 6E. As these variables tend to covary in health and in disease, Figure 6F presents simulations if we assume all tissue parameters change as a function of the input 'true' $T_{1,obs}$. A difference of 70 ms is observed between VFA and MP2RAGE corroborating the results presented in Figure 4A. MP2RAGE $T_1$ times diverge from IR with increasing $T_1$ and are 20 ms longer than IR at $T_1$ times typical for WM (~850 ms), and 70 ms longer at $T_1$ times representative of GM (1400 ms).

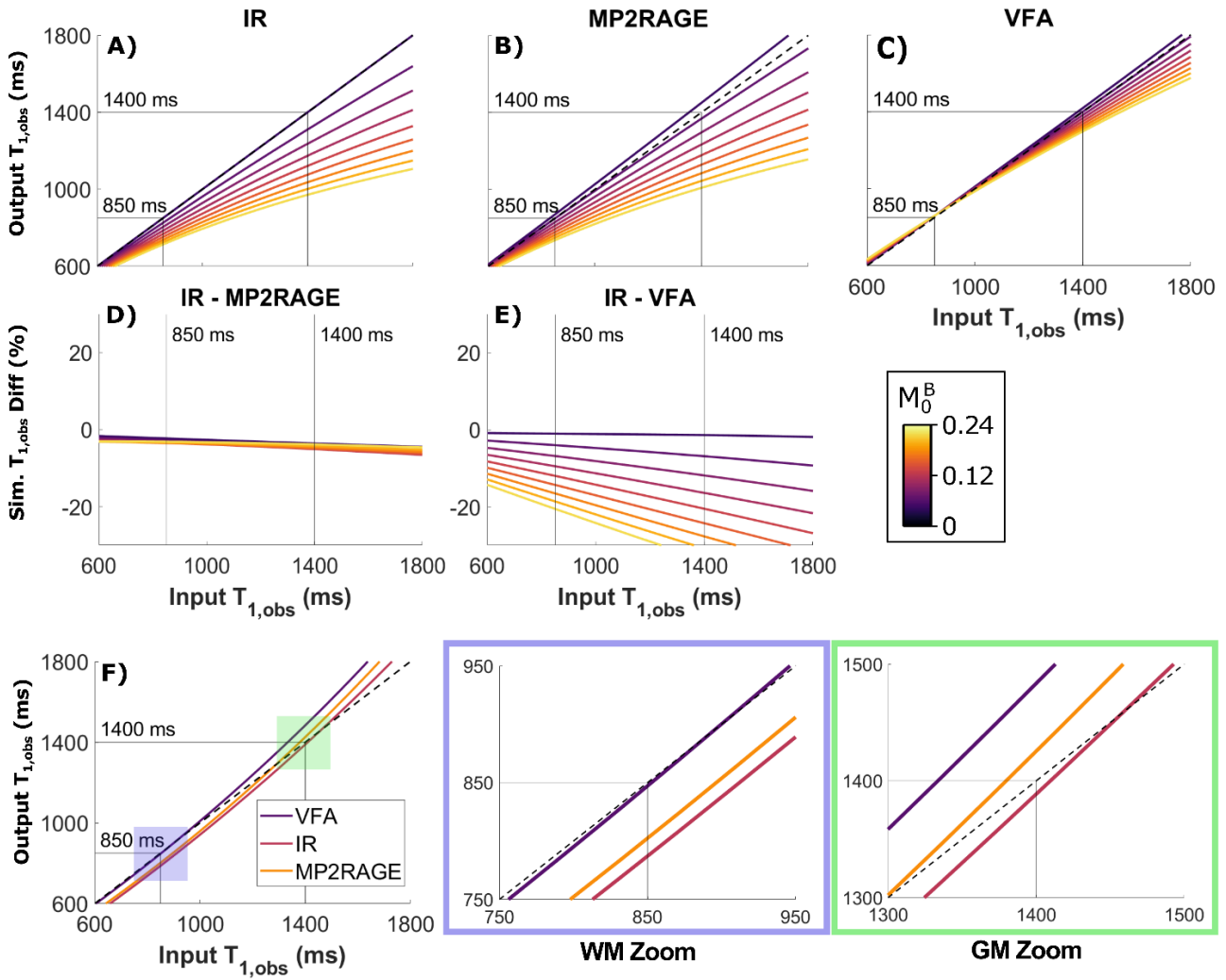

**Figure 6:** The impact of MT-effects on $T_1$ values. **A, B)** A strong MT-effect is observed for IR and MP2RAGE methods, arising from the 180-degree inversion pulse. **C)** VFA is impacted by MT-effects to a lesser degree than IR and MP2RAGE. The differences against the gold-standard IR are presented in **D)** and **E)**. **F)** In healthy tissue, multiple tissue parameters tend to change with $T_1$. By estimating tissue parameters from the qMT values obtained in (Sled and Pike, 2001), we observe a smaller impact of the MT-effect. An average of 70 ms is observed between VFA and MP2RAGE, and an additional 70 ms between MP2RAGE and IR, which converges at shorter $T_1$ times. Reference lines indicate typical $T_1$ values for GM and WM at 3T.

The csMP2RAGE $MT_{sat}$ Protocol Presents Strong Repeatability

**Figure 7** presents the test-retest $MT_{sat}$ results using csMP2RAGE for the $T_1$ mapping protocol in 10 healthy controls. We observe good test-retest values with minimal bias between scans (<1%). $T_1$ was the most repeatable parameter with an MDC of 2.95%. $MT_{sat}$ presented the largest MDC of 7.88%. A larger value is expected as the errors in the measurement of both $M_0$ and $T_1$ propagate through the calculation. If we simulate a realistic input SNR of 70 at 3T, then we expect the calculated MP2RAGE $T_1$ map to have an SNR of 55 and the $MT_{sat}$ map to have an SNR of 22 from image noise alone. This represents an SNR decrease of a factor of 2.5, which is close to the value we find *in-vivo* of 2.67, which also includes physiological noise.

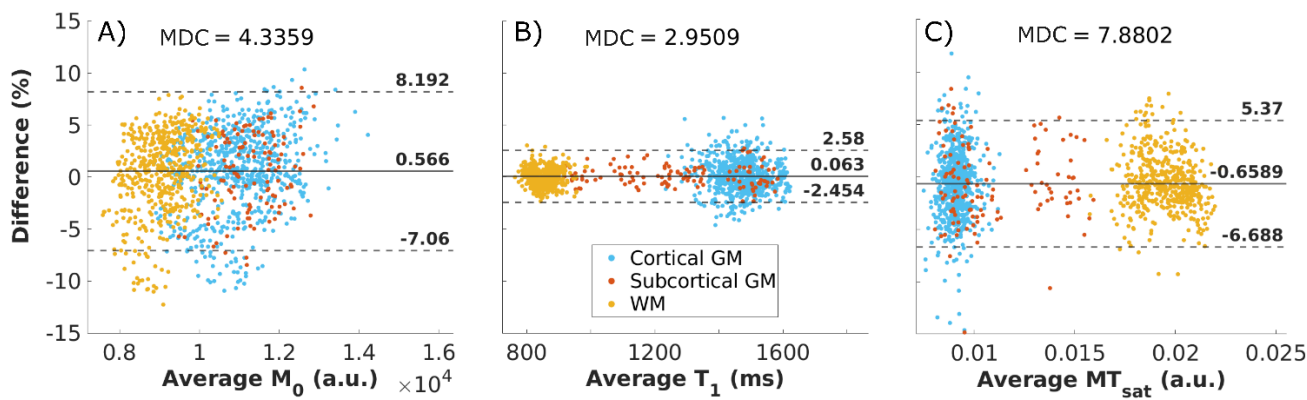

**Figure 7:** Test-retest ROI analysis using the compressed sensing based MP2RAGE for accelerated $MT_{sat}$ imaging. The solid black line indicates the average difference, and the dashed lines represent 1.96·standard deviation of the difference values, providing the 95% confidence interval. MDC = minimum detectable change.

### Discussion:

This study investigated the use of MP2RAGE $T_1$ mapping for accelerating $MT_{sat}$ imaging. The MP2RAGE $T_1$ mapping protocol reduced the scan time by 13.7% compared to the VFA protocol, with csMP2RAGE being 32.4% shorter than the VFA protocol. Using the NIST phantom, we found that all approaches could provide accurate $T_1$ values in aqueous phantoms (less than 5% deviation from reference), but the VFA $T_1$ values were more impacted by differences in $T_2$ when not correcting for incomplete spoiling. In vivo, we replicated previous reports of VFA producing larger $T_1$ values (Stikov et al., 2015b; Tsialios et al., 2017), even when correcting for incomplete spoiling and $\Delta B_1^+$. $\Delta B_1^+$ correction was shown to have a larger impact on VFA than MP2RAGE $T_1$ values, and an empirical correction produced comparable results to a model-based $\Delta B_1^+$ correction of $MT_{sat}$ values. Bloch-McConnell simulations suggest that the residual bias in the $T_1$ values between VFA and IR-based approaches (MP2RAGE and IR) could be explained by the MT-effects from the inversion pulse. Finally, we reported on the MDC from the csMP2RAGE protocol to characterize the ability of this accelerated approach to detect changes.

#### csMP2RAGE facilitates shorter $MT_{sat}$ acquisition times.

We show that the $MT_{sat}$ protocol can be accelerated using a single MP2RAGE scan instead of acquiring two images needed for VFA $T_1$ mapping, with additional time saved using csMP2RAGE. Previous work has looked to accelerate the VFA protocol with sparse sampling techniques (Berg et al., 2022; Paajanen et al., 2023; Tamada et al., 2018; Zhang et al., 2015; Zhao et al., 2015). It was shown that a 1 mm isotropic multi-echo $MT_{sat}$ protocol using a compressed sensing VFA protocol could be acquired in 15:40 (min:sec) using a 4x subsampling and in 10:30 with 6x subsampling factor (Berg et al., 2022). This produced high-quality quantitative maps where metric variance increased with acceleration factor, with minimal acceleration-induced bias (Berg et al., 2022). The total protocol using csMP2RAGE in this study was 10:46 with a 4.6x subsampling factor, with further time savings possible if compressed sensing was used in the MT-weighted image. While this work used cartesian k-space sampling, $T_1$ mapping can also be accelerated with undersampled non-cartesian trajectories combined with model-based reconstructions (Maier et al., 2019).

It is worth noting that $MT_{sat}$ imaging is typically part of a multi-parametric mapping protocol which also uses multi-echo data for $T_2^*$ fitting (Weiskopf et al., 2013). It is possible to use a multi-echo MP2RAGE sequence to obtain this information as well (Caan et al., 2019; Metere et al., 2017; Sun et al., 2020), however this study focuses on the $T_1$-portion of the protocol. In the NIST phantom, the MDC values were lowest in the VFA protocol, and averaging data across echoes could further increase the SNR and lower the MDC. This increase in SNR comes at the cost of additional scan time from the longer TR required to fit the multi-echo readout, where it may be possible to achieve similar MDC values in a time matched MP2RAGE sequence. The time savings offered with the csMP2RAGE, with minimal impact on data quality, permits the generation of $MT_{sat}$ maps in significantly less time. The csMP2RAGE will be beneficial for imaging patient populations where movement might be an issue and/or to make time to acquire complementary data.

On the accuracy of $T_1$ mapping methods

Strong agreement was observed between the different methods for measuring $T_1$ in the NIST phantom, with greater differences *in-vivo*. We replicate a previous report showing that MP2RAGE produces $T_1$ values that are closer to IR values than VFA does (Tsialios et al., 2017). With correction for flip angle inaccuracies, we achieved VFA $T_1$ values that were similar to IR values in an aqueous phantom (~4% difference), but a larger bias was found *in-vivo* that our simulations suggest is due to MT effects. As we corrected for incomplete spoiling and flip angle inaccuracies *in-vivo*, our simulations assumed perfect spoiling with no flip angle inaccuracies. This resulted in a simulated difference of 70 ms which matched our *in-vivo* results between VFA and MP2RAGE, providing further support for the differences being driven by MT.

IR is considered the gold standard approach for obtaining reference $T_1$ values. While this may be appropriate for aqueous phantoms, our simulations demonstrate the potentially large impact that the presence of a bound pool can have on the fit $T_1$ values. Exchange between the bound and free pools is noted as the source of biexponential $T_1$ recovery (Labadie et al., 2014; Rioux et al., 2016; van Gelderen et al., 2016) leading to decreased $T_1$ values from an initial faster recovery of magnetization (Mossahebi et al., 2014). Our simulations suggest that this is one reason why VFA mapping may produce elevated $T_1$ values *in-vivo* compared to the more MT-intensive IR approaches. A previous solution proposed to facilitate inter-study comparisons was to present $T_1$ values alongside the root mean square $B_1^+$ ($B_{1rms}$) of the sequence (Teixeira et al., 2019). Future work could investigate correcting IR-based $T_1$ mapping methods for MT-effects by including these into the lookup tables. However, this would require the assumption of an increasing number of tissue parameters that may or may not change concurrently. Unexpected tissue variations could lead to greater errors, which could be an important factor when using these methods to study pathologies.

The results of this study suggest that VFA and MP2RAGE can measure repeatable $T_1$ values at 3T over the expected range of $T_1$ times of brain tissue. Previous studies have investigated the repeatability of VFA $T_1$ mapping. In a similar study in the spinal cord, the MDC for VFA-derived $T_1$ was ~9% (Lévy et al., 2018). Strong repeatability and inter-site reliability values have been exhibited for VFA mapping using thick slices and four flip angles, which yielded even higher repeatability of their MTR metrics (Schwartz et al., 2019). A reproducibility study of a multi-parameter mapping protocol demonstrated lower intra- and inter-site coefficients of variation in their $R_1$ metrics compared to $MT_{sat}$ (Leutritz et al., 2020). While we found VFA-derived $T_1$ values to be most repeatable in a NIST phantom, a previous study found improved accuracy and precision when using IR over VFA, with VFA errors increasing when moving from 1.5T to 3T. Ultimately the repeatability and reproducibility will depend on the hardware and sequence timing used.

The impact of $\Delta B_1^+$ on measured $T_1$ times

It is well known that VFA $T_1$ values need to be corrected for $\Delta B_1^+$ (Boudreau et al., 2017; Stikov et al., 2015a) and for incomplete spoiling (Baudrexel et al., 2018; Corbin and Callaghan, 2021; Preibisch and Deichmann, 2009; Yarnykh, 2010). However, many variables that may be unknown to users, such as spoiling parameters, are required to perform these corrections. MP2RAGE is less sensitive to $\Delta B_1^+$ (Marques et al., 2010), but reproducibility can be improved by including $\Delta B_1^+$ in the lookup table (Haast et al., 2021). Our study reinforces these reports with substantially more variation observed between VFA $T_1$ values with and without $\Delta B_1^+$ correction compared to MP2RAGE. The effects of incomplete spoiling are likely to appear when comparing the $T_1$ of ROIs with different $T_2$ values, as spoiling phase increments are optimized for a specific $T_2$ (Yarnykh, 2010). While MP2RAGE may also suffer from incomplete spoiling issues, our results in the NIST phantom suggest that the magnitude of this effect is greater in the VFA approach. This is due to the idle time in the sequence that permits the formation of the $T_1$ contrast allowing the transverse magnetization to dissipate.

While this study is largely focused on the $T_1$ aspects of $MT_{sat}$ imaging, it should be noted that $\Delta B_1^+$ impacts the MT-weighted image and the subsequent $MT_{sat}$ value as well. Correcting for these effects will improve repeatability and reduce the MDC. This was previously addressed by using an empirical factor to correct $MT_{sat}$ values that had no corrections applied during the calculation (Helms, 2015; Weiskopf et al., 2013). These correction factors are derived by generating $MT_{sat}$ maps from images with different transmit field strengths to characterize variations as a function of $B_1^+$. Here, we demonstrate that similar results can be achieved using a model-based correction that permits separate correction of $T_1$, $M_0$, and $MT_{sat}$ values (Rowley et al., 2021). An additional benefit of correcting these effects separately is that it permits the addition of other corrections, such as for incomplete spoiling. Importantly, as shown in this study, it permits the flexibility of using different $T_1$ mapping techniques to generate $MT_{sat}$ maps that are corrected for $\Delta B_1^+$ without needing to acquire additional data to generate new empirical correction factors. This is important as the acquisition of the necessary data for generating these correction factors can require the use of custom MRI sequences.

The reduced sensitivity of MP2RAGE to $\Delta B_1^+$ effects may also be an important consideration for imaging at 7T, where these variations are larger across the brain. MP2RAGE may be a practical option at ultra-high field as MP2RAGE images are often collected in place of an MPRAGE to obtain a uniform $T_1$-weighted anatomical image. Additionally, initial work has demonstrated that MP2RAGE $T_1$ mapping is more reproducible compared to the VFA approach at 7T using single-channel transmission (sTx) (Voelker et al., 2021). It is possible to reduce $\Delta B_1^+$ using parallel transmission (pTx), however $\Delta B_1^+$ techniques would need to be adjusted to account for the effective spatial dependence of the radiofrequency pulses, particularly for MT-effects.

In conclusion, this study demonstrates that the MP2RAGE sequence can successfully replace the VFA portion of an MT$_{sat}$ protocol to remove the T$_1$ effects while significantly reducing total scan time. We have quantified the difference between the values obtained from both approaches to facilitate future comparisons across studies using different T$_1$ mapping methods.

**Acknowledgements:**

The authors thank Prof. Nikola Stikov and Agah Karakuzu (Polytechnique Montreal) for providing the NIST phantom for this study. The McConnell Brain imaging Centre is supported by Healthy Brains for Healthy Lives, the Brain Canada Foundation, through the Canada Brain Research Fund, with the financial support of Health Canada. This project was funded by the Natural Sciences and Engineering Research Council of Canada, the Fonds de recherche du Québec – Santé, the Campus Alberta Innovates Program, Quebec BioImaging Network, Jeanne Timmins Costello and Dr. David T.W. Lin Fellowships.

imaging using Cartesian phyllotaxis readout and compressed sensing reconstruction. Magn. Reson. Med. 84, 1881–1894. https://doi.org/10.1002/mrm.28244

Paajanen, A., Hanhela, M., Hänninen, N., Nykänen, O., Kolehmainen, V., Nissi, M.J., 2023. Fast Compressed Sensing of 3D Radial T1 Mapping with Different Sparse and Low-Rank Models. J. Imaging 9, 151. https://doi.org/10.3390/jimaging9080151

Preibisch, C., Deichmann, R., 2009. Influence of RF spoiling on the stability and accuracy of T1 mapping based on spoiled FLASH with varying flip angles. Magn. Reson. Med. 61, 125–135. https://doi.org/10.1002/mrm.21776

Rioux, J.A., Levesque, I.R., Rutt, B.K., 2016. Biexponential longitudinal relaxation in white matter: Characterization and impact on T1 mapping with IR-FSE and MP2RAGE. Magn. Reson. Med. 75, 2265–2277. https://doi.org/10.1002/mrm.25729

Rowley, C.D., Campbell, J.S.W., Leppert, I.R., Nelson, M.C., Pike, G.B., Tardif, C.L., 2023. Optimization of acquisition parameters for cortical inhomogeneous magnetization transfer (ihMT) imaging using a rapid gradient echo readout. Magn. Reson. Med. 1–14. https://doi.org/10.1002/mrm.29754

Rowley, C.D., Campbell, J.S.W., Wu, Z., Leppert, I.R., Rudko, D.A., Pike, G.B., Tardif, C.L., 2021. A Model-based Framework for Correcting B1+ Inhomogeneity Effects in Magnetization Transfer Saturation and Inhomogeneous Magnetization Transfer Saturation Maps. Magn. Reson. Med. 86, 2192–2207. https://doi.org/10.1002/mrm.28831

Schwartz, D.L., Tagge, I., Powers, K., Ahn, S., Bakshi, R., Calabresi, P.A., Todd Constable, R., Grinstead, J., Henry, R.G., Nair, G., Papinutto, N., Pelletier, D., Shinohara, R., Oh, J., Reich, D.S., Sicotte, N.L., Rooney, W.D., 2019. Multisite reliability and repeatability of an advanced brain MRI protocol. J. Magn. Reson. Imaging 50, 878–888. https://doi.org/10.1002/jmri.26652

Sled, J.G., Pike, B.G., 2001. Quantitative imaging of magnetization transfer exchange and relaxation properties in vivo using MRI. Magn. Reson. Med. 46, 923–931. https://doi.org/10.1002/mrm.1278

Stikov, N., Boudreau, M., Levesque, I.R., Tardif, C.L., Barral, J.K., Pike, G.B., 2015a. On the accuracy of T1 mapping: Searching for common ground. Magn. Reson. Med. 73, 514–522. https://doi.org/10.1002/mrm.25135

Stikov, N., Campbell, J.S.W., Stroh, T., Lavelée, M., Frey, S., Novek, J., Nuara, S., Ho, M.K., Bedell, B.J., Dougherty, R.F., Leppert, I.R., Boudreau, M., Narayanan, S., Duval, T., Cohen-Adad, J., Picard, P.A., Gasecka, A., Côté, D., Pike, G.B., 2015b. In vivo histology of the myelin g-ratio with magnetic resonance imaging. Neuroimage 118, 397–405. https://doi.org/10.1016/j.neuroimage.2015.05.023

Stupic, K.F., Ainslie, M., Boss, M.A., Charles, C., Dienstfrey, A.M., Evelhoch, J.L., Finn, P., Gimbutas, Z., Gunter, J.L., Hill, D.L.G., Jack, C.R., Jackson, E.F., Karaulanov, T., Keenan, K.E., Liu, G., Martin, M.N., Prasad, P. V., Rentz, N.S., Yuan, C., Russek, S.E., 2021. A standard system phantom for magnetic resonance imaging. Magn. Reson. Med. 86, 1194–1211. https://doi.org/10.1002/mrm.28779

Sun, H., Cleary, J.O., Glarin, R., Kolbe, S.C., Ordidge, R.J., Moffat, B.A., Pike, G.B., 2020. Extracting more for less: multi-echo MP2RAGE for simultaneous T1-weighted imaging, T1 mapping, R2∗ mapping, SWI, and QSM from a single acquisition. Magn. Reson. Med. 83, 1178–1191. https://doi.org/10.1002/mrm.27975
20

**Tables:**

| Parameter | VFA (PDw/T1w) | MP2RAGE | csMP2RAGE | MT-weighted GRE | $B_1^+$-map | Spin Echo IR |
|---|---|---|---|---|---|---|
| TR (ms) | 27/15 | 5000 | 5000 | 27 | 20 000 | 2000 |
| Flip Angle (degrees) | 6/20 | 4/5 | 4/5 | 6 | 8 | 90 |
| Acceleration | 6/8 Partial Fourier + GRAPPA 2 | 6/8 Partial Fourier + GRAPPA 3 | 4.6x Upsampling, density = 0.5, jitter radius = 1.2, 20 iterations, 6e-4 reg | 6/8 Partial Fourier + GRAPPA 2 | NA | None |
| TE (ms) | 2.76 | 2.76 | 2.66 | 2.76 | 2.22 | 8.5 |
| FOV | 224 x 176 | 256 x 208 | 256 x 208 | 224 x 176 | 96x96x45 | 224x224 |
| Voxel Size (mm) | 1x1x1 | 1x1x1 | 1x1x1 | 1x1x1 | 2.5x2.5x3 | 1.8x1.8x4 |
| Turbo factor | 1 | 208 | 175 | 1 | 96 | 1 |
| TI (ms) | NA | 940/2830 | 940/2830 | NA | NA | 30, 250, 500, 750, 1000, 1500 |
| MT-parameters | NA | NA | NA | 12 ms Gaussian, Δ = 2 kHz $B_{1rms}$ = 3.26 μT | NA | NA |
| Acquisition Time (min:sec) | 5:58/3:20 [total 9:18] | 7:07 | 4:18 | 5:58 | 0:40 | 4:12 |
| Total $MT_{sat}$ Protocol time | 15:56 | 13:45 | 10:46 | | | |

**Table 1: Imaging parameters**

| T₁ array (NiCl₂) | | T₂ array (MnCl₂) | |
| --- | --- | --- | --- |
| $T_1$ (ms) | $T_2$ (ms) | $T_1$ (ms) | $T_2$ (ms) |
| 1989 | 1465 | 2480 | 581.3 |
| 1454 | 1076 | 2173 | 403.5 |
| 984.1 | 717.9 | 1907 | 278.1 |
| 706 | 510.1 | 1604 | 190.94 |
| 496.7 | 359.6 | 13332 | 133.27 |
| 351.5 | 255.5 | 1044 | 96.89 |
| | | 801.7 | 64.07 |
| | | 608.6 | 46.42 |
| | | 458.4 | 31.97 |
| | | 336.5 | 22.56 |

**Table 2: NIST phantom reference values:**

**Appendix:**

<u>Linear Models for Estimating Tissue Parameters in Simulations</u>

The following equations were used for the simulations presented in Figure 6F. These are based on the values in Table 3 in (Sled and Pike, 2001). An exponential relationship was used between $M_{0B}$ and $R_{1obs}$ to prevent negative values over this range.

$$M_{0B} = 0.06732 \cdot R_{1obs}^{1.569} \tag{A1}$$

$$R_{1A} = 1.135 \cdot R_{1obs} - 0.1412 \tag{A2}$$

$$T_{2A} = -0.02915 \cdot R_{1obs} + 0.08381 \tag{A3}$$

$$T_{2B} = 2.166 \cdot 10^{-6} \cdot R_{1obs} + 8.323 \cdot 10^{-6} \tag{A4}$$

$$k_f = 2.898 \cdot R_{1obs} - 0.5081 \tag{A5}$$

$$k_r = k_f / M_{0B} \tag{A6}$$